\documentclass[final,5p,times,twocolumn]{elsarticle}

 \usepackage{graphics}

\usepackage{amssymb}
\usepackage{amsmath}

\usepackage{booktabs}
\usepackage{multirow}

\journal{NIM A  RICAP-2013}

\begin{document}

\begin{frontmatter}



\title{ Cosmic ray Spectrum, Composition, and Anisotropy Measured with IceCube}

\author[addr1]{A.~Tamburro\fnref{cor1}}
\author[]{for~the~IceCube~Collaboration\fnref{cor2}}
\address[addr1]{Bartol Research Institute and Department of Physics and Astronomy, University of Delaware, Newark, DE 19716, USA}
\fntext[cor1l]{e-mail:
atamburro@icecube.wisc.edu (Alessio Tamburro)}
\fntext[cor2]{Please refer to http://icecube.wisc.edu/collaboration/authors/2013/03 for a complete list of authors.}

\begin{abstract}
Analysis of cosmic ray surface data collected with the IceTop array of Cherenkov detectors at the South Pole provides an accurate measurement of the cosmic ray spectrum and its features in the "knee" region up to energies of about 1~EeV. IceTop is part of the IceCube Observatory that includes a deep-ice cubic kilometer detector that registers signals of penetrating muons and other particles. Surface and in-ice signals detected in coincidence provide clear insights into the nuclear composition of cosmic rays. IceCube already measured an increase of the average primary mass as a function of energy. We present preliminary results on both IceTop-only and coincident event analyses. Furthermore, we review the recent measurement of the cosmic ray anisotropy with IceCube.
\end{abstract}

\begin{keyword}
cosmic rays \sep energy spectrum \sep nuclear composition \sep anisotropy

\end{keyword}

\end{frontmatter}


\section{IceCube and Cosmic Rays}
\label{first}

Acceleration of galactic cosmic particles in the shock waves of nearby supernova remnants
is believed to produce the features observed in the energy spectrum
of primary particles detected at Earth. The gradual steepening of the cosmic ray flux at a few 10$^{15}$~eV, called the {\it knee},
and other structures at higher energies are interpreted as the signatures of these sources~\cite{Blasi:2011fi,Blasi:2011fl}.
The transition from galactic to extragalactic components of cosmic particles
is predicted~\cite{Berezinsky:2007wf} at a few 10$^{17}$~eV or 
10$^{18}$~eV depending on whether the extragalactic component is purely protonic
or a mixture of different nuclei.
At energies between 10$^{15}$~eV to 10$^{17}$~eV, 
all air shower experiments observe an increase in 
the measured average mass, compatible  
with an energy dependent change of
cosmic ray composition. Above 10$^{17}$~eV and up to 10$^{18}$~eV,
measurements of composition indicate a decrease 
of the average mass~\cite{Kampert:2012mx}.

\begin{figure}[h]
\centering
\includegraphics[width=0.5\textwidth]{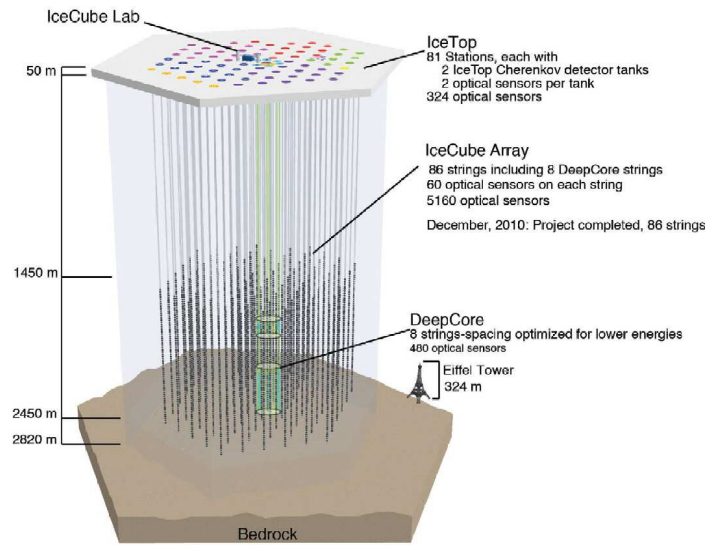}\protect\label{fig::icecube}
 \caption{
 Sketch of the IceCube Observatory.
IceCube in its 2006-07 configuration is shown in red
and referred to as IT26/IC22 (26 IceTop stations/22 in-ice cables). 
Other configurations are IT40/IC40 (2007-08) in green, 
IT59/IC59 (2008-09) in purple, IT73/IC79 (2009-10) in blue, and
IT81/IC86 (2010-11 and subsequent seasons) in yellow.}
 \end{figure}
IceCube (Fig.~\ref{fig::icecube}) is a multi-purpose astrophysical observatory installed at
the South Pole in operation since 2005~\cite{fyp}. 
It consists of a surface array of Cherenkov tanks, called IceTop, and a large
array of optical modules in the deep ice between 1.45 and 2.45 km below the ice sheet. 
Data of the surface array allow reconstructing direction and energy of down-going
primary cosmic rays in the energy range from about 100~TeV to 1~EeV. 
The main purpose of the deep detector array is to detect neutrinos,
but it also  permits the reconstruction of penetrating cosmic ray muons~\cite{ignacio}.
Since May 2011, IceCube is taking data in its full configuration.
The deep detector is an array of 86 cables (``strings''), each instrumented with 60 digital optical modules~\cite{DOM}  (DOMs).
A DOM contains photomultipliers and readout electronics. 

\begin{figure}[h]
    \centering
   \includegraphics[width=0.45\textwidth]{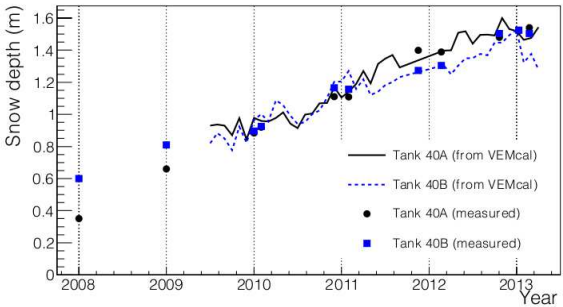}
    \caption{\label{fig::snow}{ Accumulation of snow over time,
for the two tanks of one IceTop station~\cite{snow}. Direct
measurements of snow depth are shown with solid symbols.
Indirect measurements using calibration curves (VEMCal data)
are shown with lines.}}
 \end{figure}

Near the top of each string at an altitude of 2835~m a.s.l. (atmospheric depth 
of about 680~g/cm$^2$), 81 pairs (``stations'') of cylindrical Cherenkov tanks
form IceTop, covering an area of about 1~km$^2$~\cite{IT}.
Each tank contains two standard IceCube DOMs	
and samples secondary particles (low-energy photons, electrons, and muons) from air showers.
At the time of deployment, the top of each tank was at the same level
as the surrounding snow. However, snow drifting causes the overburden to increase with time.
The snow depth over each tank is physically measured every year 
and can be indirectly estimated from the muon/electron ratio in calibration curves (Fig.~\ref{fig::snow}). 
IceTop DOM charges are calibrated using signals
from single muons, expressed as an
independent unit called ``Vertical Equivalent Muon'' (VEM)~\cite{IT}.

\begin{figure}[t]
\centering
\includegraphics[width=0.2\textwidth]{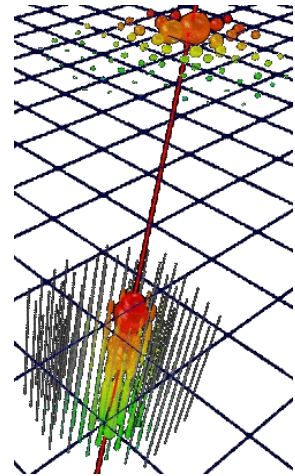}
\caption{\protect\label{fig::coinceve}
Cosmic ray coincident event recorded by IceCube in 2010. Triggered
DOMs are indicated with colored spheres whose volume is proportional to the registered signal.
Signal times are indicated with colors from red to green, red being the earliest. This event
is reconstructed with an energy of 3$\cdot$10$^{17}$~eV and
a zenith angle of 11.5$^{\circ}$. Additional information on surface and in-ice reconstruction
is given in Figs.~\ref{fig::dlp} and \ref{fig::dedx}, Sec.~\ref{second} and \ref{third}.}
\end{figure}

At the altitude of IceTop, secondary particles are sampled near the shower maximum.
This allows precisely measuring the energy
spectrum of primary cosmic rays with an energy resolution of about 10\% above 10$^{16}$~eV (10~PeV).
Events seen in coincidence by both IceTop and the deep detector 
(see Fig.~\ref{fig::coinceve}) give clear insights 
into the nuclear composition of cosmic rays
for energies that span from PeV to EeV (10$^{18}$~eV).
The deep detector measures the signal of penetrating muons
(more than about 500~GeV at production) from the early stage of shower development.
The in-ice DOMs detect the light emitted due to 
energy loss of high energy muons inside the detector volume.
The amount of Cherenkov light generated is proportional to
the deposited energy, which is in good approximation a function of the
muon multiplicity alone. 
At a fixed time, the light collected from these muons has traveled a certain distance and 
the coherent wave front of photons emitted at a given time
is used to reconstruct track direction and energy loss profile.
Penetrating muons are more abundant in iron showers than in proton showers since 
shower development starts higher in the atmosphere.
The in-ice signal of iron showers is therefore larger for a given energy and zenith angle.

During the construction phase, IceTop measured the cosmic ray energy spectrum at energies 
between 1~PeV and 100~PeV when only 26 stations were operational~\cite{it26}.
Using coincident events, the cosmic ray spectrum and 
average nuclear composition were measured between 1~PeV and 30~PeV~\cite{ic40}.
Furthermore, the large statistics and good
angular resolution allowed detection of cosmic ray anisotropies in the Southern sky at the
{\it per-mille} level on angular scales down to a few degrees~\cite{Abbasi:2010mf,Abbasi:2011zka}.

This paper emphasizes and highlights the latest results from analyses of cosmic ray events
collected with 73 stations of IceTop (IT73) and 79 cables of IceCube (IC79). 
The latest measurements of energy spectrum and composition 
cover the energy range from about 1~PeV to 1~EeV and
include zenith angles up to about 40$^\circ$. Only events with reconstructed
shower cores contained within the IceTop array were considered.
In Sec.~\ref{second} the primary energy spectrum obtained
with events of IT73-only is presented. In Sec.~\ref{third}
the measurement of composition with coincident events of IC79/IT73
is discussed. In Sec.~\ref{fourth} a recent study of the anisotropy comparing 
IceTop and deep detector measurements is reviewed. 
An outlook on the status of the extension of the current analysis to include more inclined events
is given in Sec.~\ref{fifth}.

\section{Cosmic ray Primary Spectrum with IceTop}
\label{second}

Using data taken between June 1, 2010 and May 13, 2011
(effective livetime of 327 days), about 37 million events of cosmic rays were reconstructed. 
These events are a selection of IceTop events with $\cos\theta>$~0.8 ($\theta<$~37$^\circ$) and
triggering 5 or more stations. For the analysis, only ``contained'' events were considered.
These events had reconstructed cores within an area delimited by the outermost stations. 

\begin{figure}[h]
    \centering
   \includegraphics[width=0.45\textwidth]{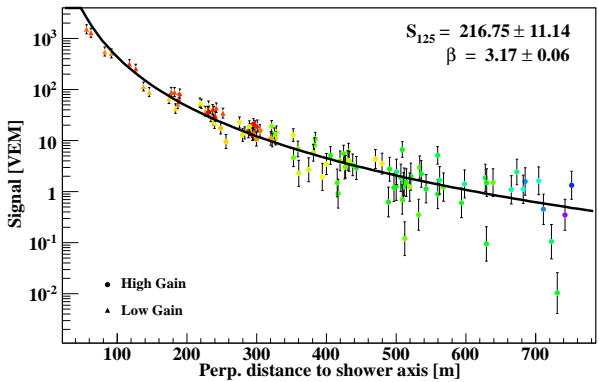}
    \caption{\label{fig::dlp}{Lateral distribution of tank
 signals in VEM for the event in Fig.~\ref{fig::coinceve} fitted to the function in Eq.~\ref{eq::dlp}.
 Below 100~m, saturation starts becoming evident with signals around 1500~VEM,
 but it does not impact significantly $S_{125}$.
 DOMs are completely saturated at about 3000~VEM.}}
 \end{figure}

The surface shower particle density decreases rapidly with the distance from
the shower axis. This lateral distribution function (LDF)
carries information about the energy of the primary particle.
The charge expectation value $S$ in an IceTop tank at distance $r$ from the shower axis is described 
by a ``double logarithmic parabola''~\cite{IT}
\begin{equation}
\label{eq::dlp}
S(r) = S_{125}\cdot \left(\frac{r}{125}\right)^{-\beta-\kappa\log_{10}(r/125)},
\end{equation}
where $S_{125}$ is the charge in VEM at 125~m (Fig.~\ref{fig::dlp}).
This description is empirical and derived from simulation. 
In log-log format, $\beta$ represents the slope of $\log_{10}S(r)$ at 125~m
and $\kappa$ represents its curvature. Assuming a fixed value of
$\kappa=$~0.303 was verified not to impair reconstruction quality of simulated events.
Signals measured between about 30~m and 300~m from the shower axis are well
described by Eq.~\ref{eq::dlp} for primary zenith angles in the range
0$^\circ$--40$^\circ$. For larger angles, the LDF needs to include
a description of the muon component as a function of the distance from the shower core (see Sec.~\ref{fifth})
The reference distance of 125~m 
makes $S_{125}$ approximately independent of the primary type and 
minimizes the correlation between $S_{125}$ and $\beta$.
Events with $\log_{10}(S_{125})\leq$~0 are currently
not used in any analysis since they are affected by fluctuations
introduced when correcting the signals to account for the snow coverage. 
Correct treatment of events below this threshold is under investigation.

Snow coverage affects the measurement of $S_{125}$.  
Expected tank signals are therefore corrected by using the following equation~\cite{snow}
\begin{equation}
\label{eq::snow}
S_{corr}(r) = S(r)\cdot exp(-X/\lambda_{s}),
\end{equation}
where $X=d_{snow}/\cos\theta$ is the slant depth that particles must traverse
to the tank at a depth $d_{snow}$, and $\lambda_s$ is an attenuation length 
fixed to 2.1~m. A maximum likelihood method is used to derive $S_{125}$ and
zenith angle from integrated charges and leading edge times of waveforms~\cite{IT}. This likelihood includes an additional
term describing signal saturation (see Fig.~\ref{fig::dlp}).

\begin{figure}[h]
\centering
\includegraphics[width=0.45\textwidth]{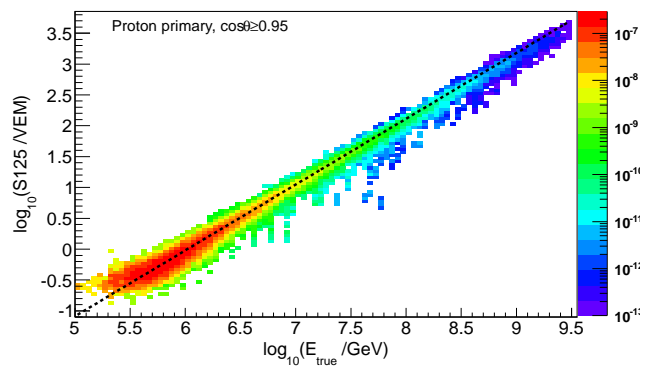}
\caption{\protect\label{fig::s123etrue}
$\log_{10}S_{125}$ $vs$ $\log_{10}E_{true}$ for simulated events of proton primaries with 
$\cos\theta>$0.95 and triggering 5 or more stations. The histogram is weighted
with $E^{-2.7}$. For a measured $\log_{10}S_{125}$, the dashed line
can be used to identify the most likely true primary energy (see text for details).}
\end{figure}

The energy of the primary cosmic ray is estimated from
the relationship between $S_{125}$ and the true primary
energy, $E_{true}$, derived from simulations. 
The 2-dimensional histogram of $\log_{10}S_{125}$ $vs$ $\log_{10}E_{true}$
for simulated proton showers with $\cos\theta>0.95$ is shown in Fig.~\ref{fig::s123etrue}.
The relationship between $\log_{10}S_{125}$ and $\log_{10}E_{true}$ depends
on the mass of the primary particle and the zenith angle of its arrival direction.
For any measured $\log_{10}S_{125}$ within a bin of size $\Delta\log_{10}S_{125}=$~0.05, 
the reconstructed energy ($\log_{10}E$) is the mean of the fitted
distribution of the corresponding $\log_{10}E_{true}$ values.
The impact of a change of about 1 unit in the spectral index (from -2.6 to -3.5) used to weight the 2-dimensional histogram of
$\log_{10}S_{125}$ $vs$ $\log_{10}E_{true}$ is negligible and corresponds
to a shift of the mean of $\log_{10}E_{true}$ distributions which is smaller than $\Delta\log_{10}E_{true}=$~0.05.
Consequently, the relationship between primary and reconstructed energy 
is not spectrum-dependent.

\begin{figure}[h]
\centering
\includegraphics[width=0.48\textwidth]{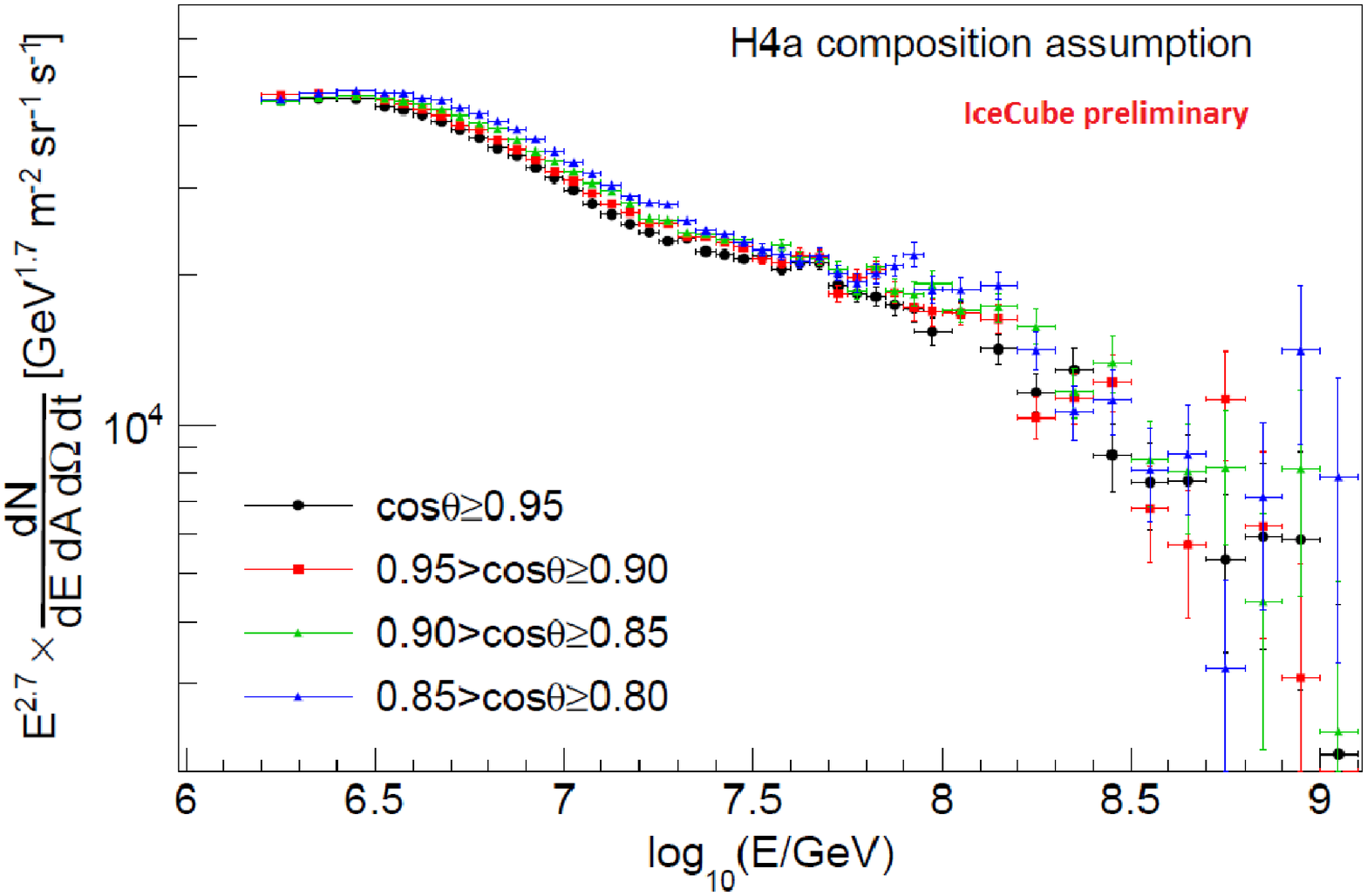}
\caption{\protect\label{fig::h4aspectrum}
IT73 cosmic ray spectrum in four different zenith angle bins assuming a primary composition based on the H4a (see text for detail).}
\end{figure}

Because of mass-dependent energy reconstruction biases, 
to derive the all-particle energy spectrum, a specific assumption
for the primary composition of the detected cosmic ray events has to be made. 
Assuming an isotropic flux of cosmic rays, one expects that the energy spectrum is invariant with respect
to the zenith angle. Energy reconstruction biases do depend on the zenith angle,
therefore an incorrect composition assumption leads to a misalignment
between measurements at different angles.
In this paper and in Ref.~\cite{it73_1} the final result is given using the composition assumption 
corresponding to the $H4a$  model described in Ref.~\cite{Gaisser:2012zz}. 
The choice of this specific model is justified
by a reasonable agreement of the measured angular spectra 
in different energy bands (Fig.~\ref{fig::h4aspectrum}).

To unfold the final spectrum and determine its shape, an iterative procedure was used.
At each step the spectrum is evaluated based on
the effective area and the relationship
$\log_{10}S_{125}$ $vs$ $\log_{10}E_{true}$
from the previous step.
The fractional contributions of the elemental groups of the H4a model
was kept throughout the procedure.
Starting with a featureless power law spectrum,
the final spectrum was obtained after two iterations.

\begin{figure}[h]
\centering
\includegraphics[width=0.45\textwidth]{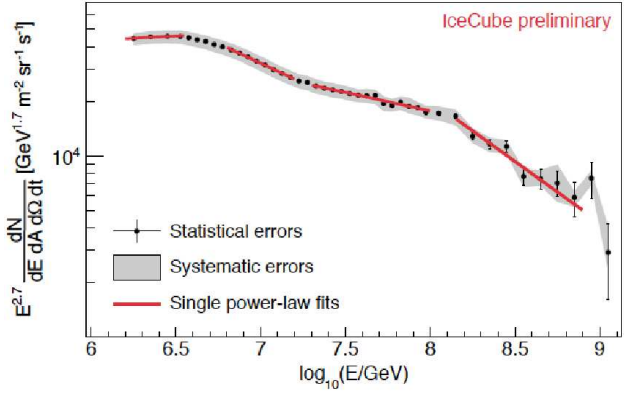}
\caption{\protect\label{fig::h4asyst}
Cosmic ray energy spectrum assuming a composition corresponding to the H4a model.
Systematic and statistical
uncertainties are shown along with spectral fits in four different energy ranges (see Tabs.~\ref{tab::ind},\ref{tab::syst}  ).}
\end{figure}

In Fig.~\ref{fig::h4asyst} the measured all-particle energy spectrum
is shown with its systematic and statistical uncertainties. 
The all-particle energy spectrum does not
follow a single power law above the knee measured at about 4.4~PeV.
The spectrum was fitted by a power law function proportional to
$E^{-\gamma+1}$ in four different energy ranges.
The selected ranges did not include data points where
the transition between two power laws was observed.

The spectral index $\gamma$ for the four energy ranges is listed
in Tab.~\ref{tab::ind}. 
\begin{table}[h]
\footnotesize
 \caption[]{\label{tab::ind} Spectral indices with statistical and systematic uncertainties
 in the four energy ranges of $\log_{10}\left(E/GeV\right)$ shown in Fig.~\ref{fig::h4asyst}.}
 \begin{center}
   \begin{tabular}{ccc} \toprule[2pt]
     Energy range & $\gamma\pm$stat.$\pm$sys. & $\chi^2/ndf$\\
     \midrule[1pt]
      6.20--6.55 & 2.648$\pm$0.002$\pm$0.06 & 206/2\\
      6.80--7.20 & 3.138$\pm$0.006$\pm$0.03 & 14/6\\
      7.30--8.00 & 2.903$\pm$0.010$\pm$0.03 & 19/12\\
      8.15--8.90 & 3.374$\pm$0.069$\pm$0.08 & 8/6\\
       \bottomrule[2pt]			
    \end{tabular}
  \end{center}
\end{table}
Above about 18~PeV
the spectrum hardens. A sharp drop is observed
beyond 130~PeV.
The four major systematic uncertainties at two measured energies
are summarized in Tab.~\ref{tab::syst}, with the composition assumption being
the largest. Two hadronic interaction models
were used in simulations: SYBILL2.1~\cite{sybill} and QGSJETII~\cite{qgsjet}.
\begin{table}[h]
\footnotesize
 \caption[]{\label{tab::syst} Systematic uncertainties of the measured flux 
 at two energies. The flux assumes a composition corresponding to the H4a model.
The systematic uncertainties include VEM calibration, snow correction, interaction model,
 and composition assumption.}
 \begin{center}
   \begin{tabular}{cccccc} \toprule[2pt]
     \multirow{2}*{Energy} &   \multicolumn{4}{c}{Flux Sys.} &  \multirow{2}*{Total}\\
                                                      &  VEM & Snow & Inter. & Comp. & \\
     \midrule[1pt]
      3 PeV &  +4.0\% -4.2\% & +4.6\% -3.6\% & -4.4\% & $\pm$7.0\% & +9.3\% -9.9\%\\
	30 PeV & +5.3\% -5.3\% & +6.3\% -4.9\% & -2.0\% & $\pm$7.0\% & +11.0\% -10.0\\
       \bottomrule[2pt]			
    \end{tabular}
  \end{center}
\end{table}
For the same primary energy QGSJETII produced
larger $S_{125}$ signals compared to SYBILL 2.1. 
To estimate the systematics due to composition,
the differences between the final and most vertical spectra
and the final and most inclined spectra in the primary 
energy range between $\log_{10}(E/GeV)=$~6.2
and $\log_{10}(E/GeV)=$~7.5 were used. This energy range
has negligible statistical fluctuations.   
The total uncertainty on the primary flux is about 10\% in all energy ranges.

\section{Nuclear Composition of Cosmic Rays with IceCube}
\label{third}

Cosmic rays interacting in the atmosphere produce muons.
If these muons have enough energy to reach the array of in-ice DOMs of IceCube,
they can be used to probe earlier stages of the shower development.
For example, depending on the type of the primary particle, proton or iron,
an event of 5$\cdot$10$^{15}$~eV is expected to produce 30 to 80 muons with
sufficient energy to reach  a depth of 1500~m. The energy deposited in the deep
detector consequently varies from 5$\cdot$10$^{12}$~eV 
to 15$\cdot$10$^{12}$~eV.  
The composition-dependence of the muon multiplicity is the main detection principle
used for coincident event analyses.
For TeV muons, IceCube mostly detects the Cherenkov light emitted by secondary particles produced in radiative
loss processes. Individual catastrophic energy losses can be identified as abrupt
increases in the collected light.

\begin{figure}[h]
    \centering
   \includegraphics[width=0.42\textwidth]{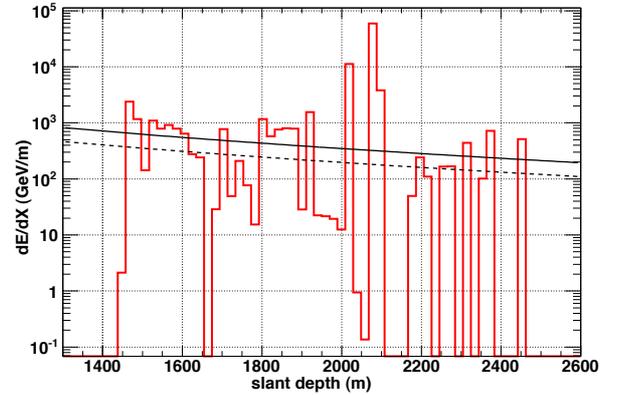}
   \vspace{-0.5cm}
    \caption{\protect\label{fig::dedx}{Reconstructed muon bundle energy loss $\left(dE/dX\right)_{bundle}$ 
    as a function of the in-ice depth for the event in Fig.~\ref{fig::coinceve}. 
     The solid black line is the fitted average energy
 loss. The dashed line is the average energy
 loss after removal of stochastic peaks (see text for detail).
 The highest peak, located at a
depth with exceptionally strong dust concentration, and bins with zero
energy loss are artifacts of the reconstruction algorithm.}}
 \end{figure}

Nearly vertical ($\theta\lesssim$~37$^\circ$ or $\cos\theta >$~0.8) 
contained air shower events of IC79/IT73 were analyzed
to study the mass composition of cosmic rays~\cite{nn}.
The subsample (30\%) of the events described in Sec.~\ref{second} 
and contained in both surface and deep detector were reconstructed
with their axis crossing both arrays of IceCube.
The energy loss profile of a muon bundle as a function of in-ice depth
$\left(dE_\mu/dX\right)_{bundle}\left(X\right)$ is derived from amplitude and timing of
the signals in the deep optical sensors. The reconstruction takes into account
absorption and scattering properties of the surrounding ice. 
An example is shown in Fig.~\ref{fig::dedx}.
The energy loss at a fixed depth depends mainly
on the muon multiplicity, thus being a composition sensitive observable. 

Charged mesons, mainly kaons and pions, either decay into high-energy
muons or re-interact, depending on the local air density. A correction
procedure is therefore applied to account for the strong seasonal
variations of the Antarctic atmosphere~\cite{seas}. 

The energy loss of a single muon is given by
\begin{equation}
\label{eq::eloss}
\frac{dE_\mu}{dX} = -b \left(E_{\mu, surf.} + \frac{a}{b}\right)\cdot e^{-bX},
\end{equation}
where $a=$~0.260~GeV/m is the continuous energy loss
constant and $b=$~3.57$\cdot$10$^{-4}$~m$^{-1}$ the proportional energy
loss constant.
The profile of a bundle produced by a shower of a primary
particle is described with the following equation~\cite{dedx}
\begin{equation}
\label{eq::dedx}
\left(\frac{dE_\mu}{dX}\right)_{bundle}\left(X\right) = \int_{E_{\mu,min}}^{E_{\mu,max}}\frac{dN_\mu}{dE_\mu} \frac{dE_\mu}{dX} dE_{\mu,surf.},
\end{equation}
where $dN_\mu/dE$ is the energy distribution of the muons in the bundle,
$E_{\mu,min} = \frac{a}{b} \left(e^{bX} - 1\right)$ the minimum energy a muon needs at the surface to reach the depth $X$, 
and $E_{\mu,max}\propto E/A$  the maximum energy a muon can obtain from this shower 
at the surface. Here, $E$ represents the primary energy and $A$ its atomic mass.
The muon multiplicity is given by the formula~\cite{elbert}
\begin{equation}
\label{eq::elbert}
\frac{dN_\mu}{dE_\mu}  = \gamma_\mu\kappa(A) \left(\frac{E}{A}\right)^{\gamma_\mu -1} E_\mu ^ {-\gamma_\mu - 1},
\end{equation}
where $\gamma_\mu=1.757$ is the muon integral spectral index
and $\kappa$ a normalization factor that depends on the shower properties.
By integrating Eq.~\ref{eq::dedx}, an average muon bundle energy loss is obtained
\begin{equation}
\label{eq::avgdedx1}
\left(\frac{dE_\mu}{dX}\right)_{bundle}\left(X\right)  =  
\kappa\cdot \left(\frac{A}{\cos\theta}\right)\cdot e^{-bX}\cdot \gamma_\mu \cdot \left(\frac{E}{A}\right)^{\gamma_\mu -1} \cdot \nonumber
\end{equation}
\begin{equation}
\label{eq::avgdedx2}
\cdot \left[-\left(\frac{E}{A}\right)^{-\gamma_\mu} \cdot \left(\frac{a}{\gamma_\mu} - \frac{b}{1 - \gamma_\mu} \cdot \frac{E}{A} \right)
+ E_{min}^{-\gamma_\mu} \cdot \left(\frac{a}{\gamma_\mu} - \frac{b}{1-\gamma_\mu} \cdot E_{min}\right) \right] . 
\end{equation}
The reconstructed profile is fitted to this integrated energy loss profile.

\begin{figure}[h]
\centering
\includegraphics[width=0.45\textwidth]{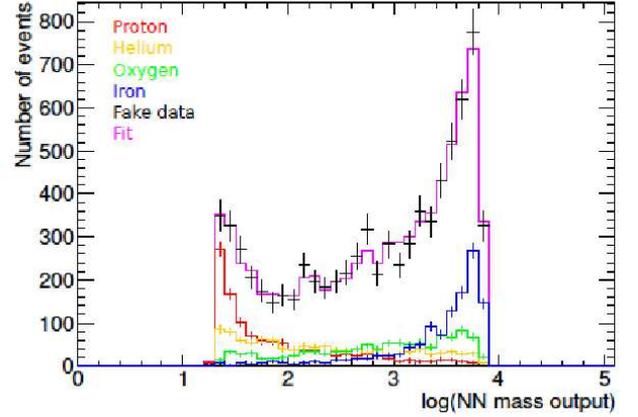}
\caption{\protect\label{fig::template}
Template histograms for four mass groups in the
reconstructed energy bin $\log_{10}E= \left[7.6,7.7\right]$ for a surrogate
dataset created from simulations.}
\end{figure}

A multilayer perceptron neural network (NN) was trained with Monte Carlo simulations
of 4 primaries (proton, helium, oxygen, and iron) to solve the non-linear mapping
of primary energy, primary mass, and reconstructed variables.
The masses were chosen due to their equal distance on a logarithmic
scale. The previous analysis of coincident events
showed that Si and Fe yield similar outputs~\cite{ic40}.
Measurements of the electromagnetic
component of the air showers at the surface ($S_{125}$) and the muon component 
in the ice ($\log_{10}\left(dE/dX\right)_{bundle}$) are the main ingredients
used to train the network how to find the best fit to the primary energy and mass.
More detail on the analysis can be found
in Ref.~\cite{nn}. Intrinsic shower fluctuations and the relatively large overlap between different
primary types in the input variables cause a considerable spread of the NN mass output
for a given reconstructed energy bin. An un-binned likelihood fit is used
to create template histograms for each simulated mass group and
reconstructed energy (Fig.~\ref{fig::template}).

\begin{figure}[h]
\centering
\includegraphics[width=0.48\textwidth]{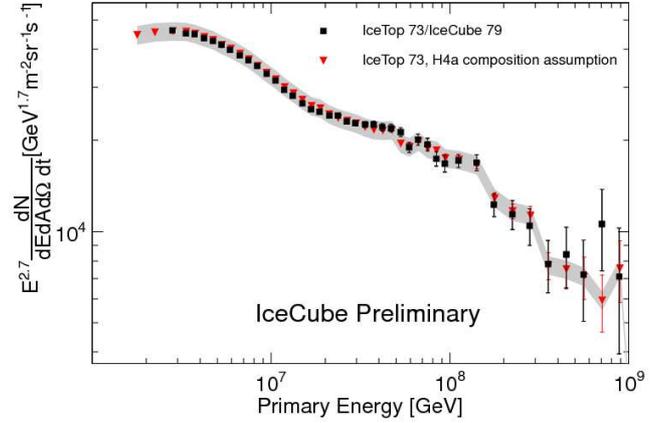}
\caption{\protect\label{fig::it73ic79comparison}
Differential energy spectrum of IC79/IT73 events multiplied by
$E^{2.7}$ compared with the IT73-only
measurement shown in Fig.~\ref{fig::h4asyst}.
A systematic error band of $\pm7\%$ 
due to the composition uncertainty is added to the IT73 spectrum.}
\end{figure}

Data were passed through the trained NN to obtain a composition-independent energy
estimate. The resulting spectrum can be compared to that derived from
IceTop-only data (Fig.~\ref{fig::it73ic79comparison}).
The coincident event analysis is independent of composition
assumptions and the good agreement with the energy spectrum obtained in Sec.~\ref{second}
demonstrates that selecting the H4a model was a reasonable choice.

For a given reconstructed energy bin, the fractions of 
each individual mass group
(p$_H$, p$_{He}$, p$_O$, p$_{Fe}$) and their uncertainties are known
from the template histogram for that bin.
\begin{figure}[h]
\centering
\includegraphics[width=0.48\textwidth]{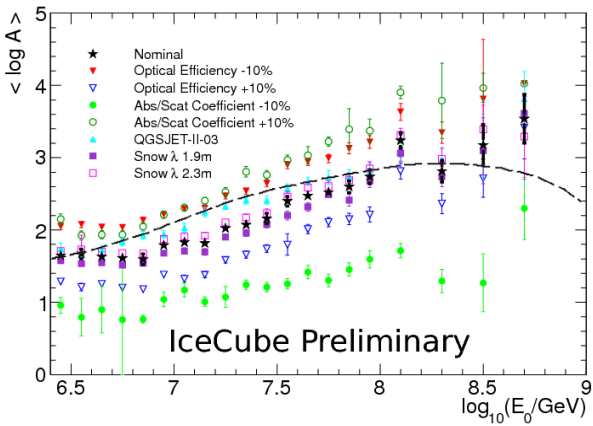}
\caption{\protect\label{fig::it73ic79comp}
The $<logA>$ composition spectrum as a function
of the primary energy obtained with IC79/IT73 data. 
Systematics are included, dashed line represents the prediction
based on the H4a model.
This figure is adapted from Ref.~\cite{nn}.}
\end{figure}
The average mass in the given energy bin is calculated as follows: 
\begin{equation}
\label{lna}
<logA> = p_H logA_H+ p_{He}logA_{He} + p_OlogA_O + p_{Fe}logA_{Fe}.
\end{equation}
The result is shown in Fig.~\ref{fig::it73ic79comp}. Above about 300~PeV 
not enough reconstructed events are available from 1 year of data.
This results in a larger uncertainty on the solution of template histograms
at these energies.
 
\section{Studies of Anisotropy with IceCube}
\label{fourth}

\begin{figure}[h]
      \centering
   \includegraphics[width=0.45\textwidth]{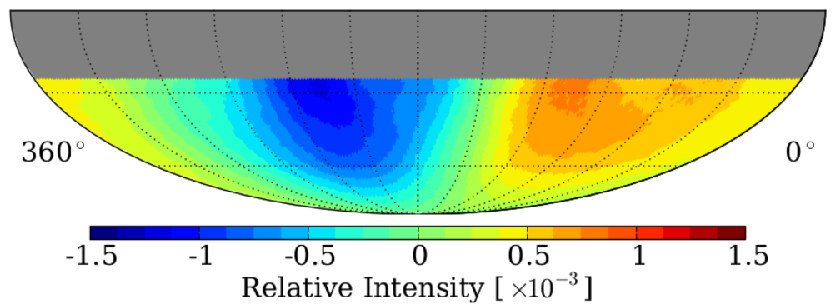}
    \includegraphics[width=0.45\textwidth]{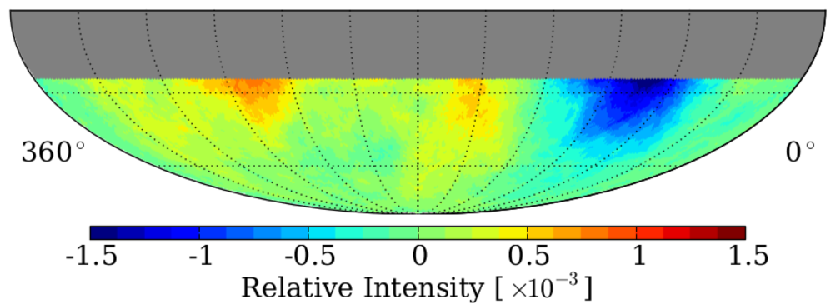}
 \caption{\protect\label{fig::icskymaps}{Relative intensity sky maps obtained 
with  IC79. The data sets have median primary energies 
of 20~TeV (upper) and 400~TeV (lower). 
The angular binning  or smoothing angle is 20$^\circ$.
The ``deepest'' deficits in the lower map is measured with a significance of 6.3$\sigma$.}}
\end{figure}

The anisotropy of cosmic ray arrival directions can be measured
in two different ways in IceCube~\cite{ani}: 
using TeV muon events collected with the deep detector~\cite{Abbasi:2010mf,Abbasi:2011zka}
and using air showers triggering the surface array.
The in-ice detector has a lower primary energy threshold than IceTop 
(20~TeV) and can use events at larger zenith angles (up to 70$^\circ$).
This makes it possible for the deep detector to
reach a higher sensitivity (about 6.3$\cdot$10$^{10}$~events/yr, 
anisotropy level $\delta >$~10$^{-5}$) and scan small scale structures.
In Fig.~\ref{fig::icskymaps} the sky maps in two energy ranges 
obtained with data collected from IceCube between May 2009 and May 2010 are shown.

\begin{figure}[h]
    \centering
\includegraphics[width=0.45\textwidth]{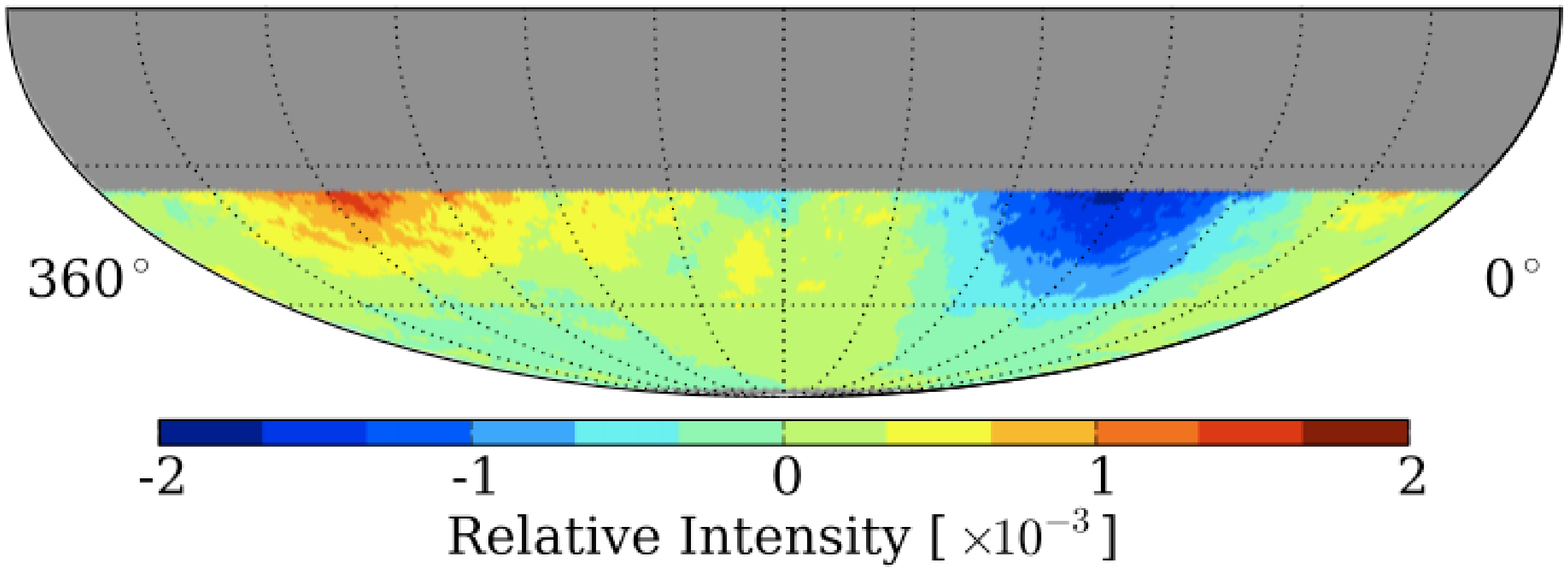}\\
\includegraphics[width=0.45\textwidth]{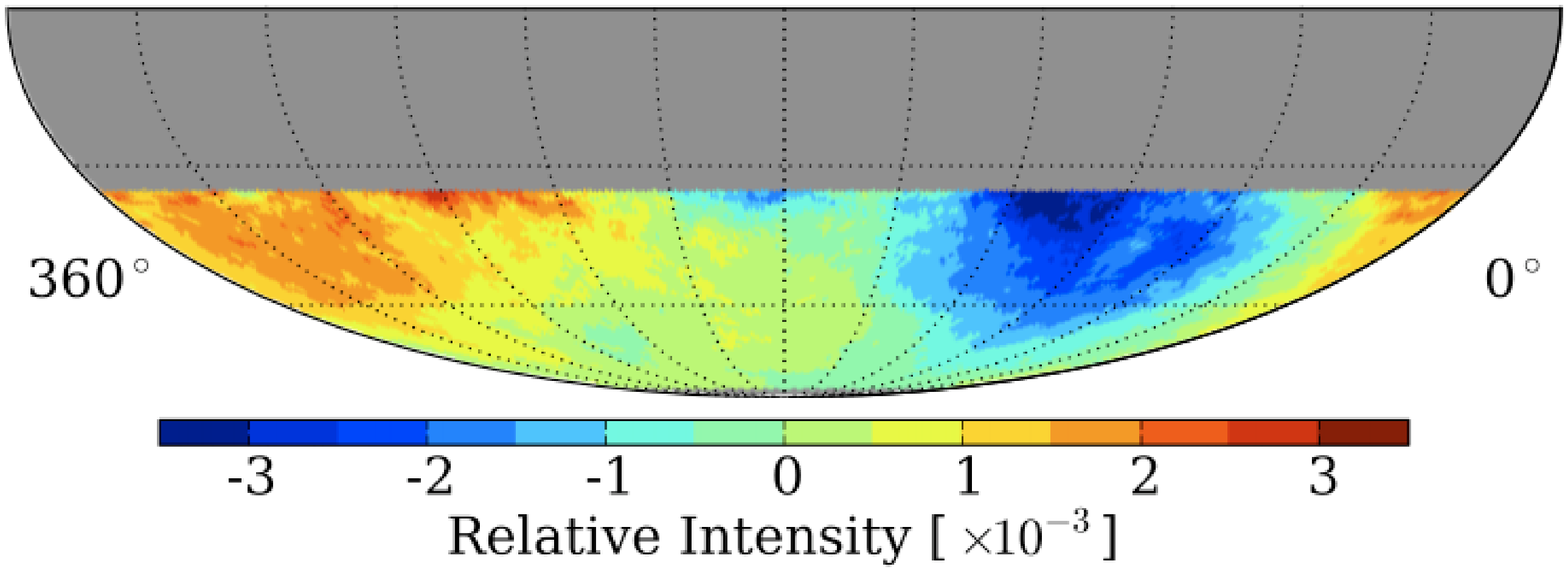}
 \caption{\protect\label{fig::skymaps}{Relative intensity sky maps obtained 
with  IT73 data. 
The IceTop datasets have a median primary energy of 
400~TeV (upper) and 2~PeV (lower). The angular binning or smoothing angle is 20$^\circ$.
For the low energy sky map, 2.9$\cdot$10$^8$ events and
for the high energy sky map, 0.7$\cdot$10$^8$ events were used.
The ``deepest'' deficits in both maps are measured with a significance of 7.1$\sigma$.}}
\end{figure}

The main advantage of IceTop is a better energy resolution (20\% at $>$300~TeV).
Fig.~\ref{fig::skymaps} shows the sky maps in two energy ranges 
obtained with IceTop data collected between May 2009 and May 2011~\cite{itani}.
Only events with zenith angles less than 60$^\circ$ are
selected (1.4$\cdot$10$^8$~events/yr above 100 TeV, sensitivity to
anisotropies $\delta >$~10$^{-4}$).
IceTop confirms the large scale anisotropy measured with deep-ice events
of IceCube. 
The result also shows that the global topology of this anisotropy does not change at
higher energies.
The measurement is inconsistent in both amplitude and phase with the Compton-Getting prediction
of an apparent anisotropy caused by
the relative motion between the Earth and sources of cosmic rays~\cite{CG}.
A detailed review of possible explanations of the observed topology can be found in Ref.~\cite{ani}.

\section{Outlook}
\label{fifth}

In Sec.~\ref{second} and \ref{third} results of analysis of
nearly vertical events have been presented. 
In particular, IceTop-only events up to about 40$^\circ$ 
and coincident IceCube events up to about 30$^\circ$
have been used to scan primary energies above about 1~PeV.
Analysis of data collected with a denser sub-array at the center of IceTop 
will decrease the energy threshold of the detector to about 100~TeV~\cite{infill}.

\begin{figure}[h]
    \centering
\includegraphics[width=0.45\textwidth]{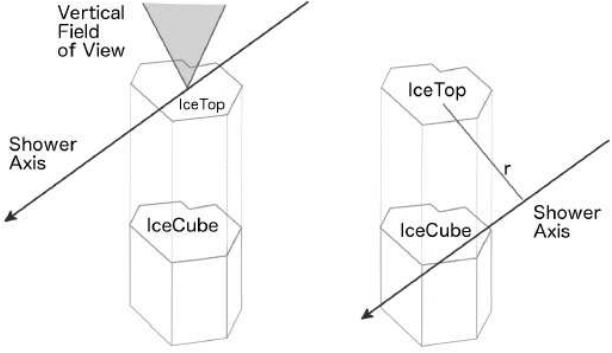}
 \caption{\protect\label{fig::inclined}{(Left) IceTop-contained
showers. The shower axis passes through the IceTop
array and misses the in-ice detector. (Right) IceCube-contained showers.
The shower axis passes through the in-ice component of IceCube.}}
\end{figure}

\begin{figure}[h]
    \centering
\includegraphics[width=0.45\textwidth]{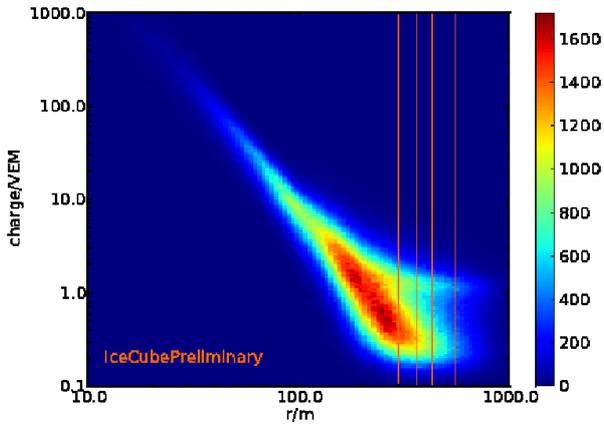}
 \caption{\protect\label{fig::cinclined}{Two-dimensional histogram 
of the tank total charge as a function of the distance to the shower
axis for events with S$_{125}$ between 4 and 5 VEM and zenith angles
between 30$^\circ$ and 33$^\circ$. At larger distances from the core (300 m and above)
the contribution from single muons becomes apparent as charge clustering around 1 VEM.}}
\end{figure}

Another ongoing analysis is trying to identify and reconstruct
inclined events up to a zenith angle of 60$^\circ$~\cite{inclined}.
Depending on the primary energy, inclined events above 30$^\circ$
and with their reconstructed shower axis traversing only the in-ice detector volume
can leave signals in IceTop tanks.
These signals have a larger contribution from the muon component of air showers.
Also, inclined events with axis intersecting IceTop but missing the in-ice volume
are expected to have a different LDF than the one described here. In fact, the larger amount of matter
encountered by secondaries causes their rapid absorption.
The use of events contained in only one of the detector components for
coincident analyses (Fig.~\ref{fig::inclined}) will more than double the current aperture.
For inclined showers, the larger muon contribution (see Fig.~\ref{fig::cinclined}) needs
a modification of the current IceTop LDF, which is optimized for nearly vertical showers
whose surface component is mainly dominated by electromagnetic particles.
The number of muons 
collected is described by:
\begin{equation}
\label{mld}
N_\mu(r) = A S_{125}^\beta r^{-0.75} \left( 1+\frac{r}{320 m}\right)^\gamma
\end{equation}
where $\beta$ and $\gamma$ are parameters that depend on energy
and direction of the shower.
Knowing the expected number of muons for tanks at large distances
from the shower core is the first step towards an energy estimate
for inclined showers. 
  
\section{Conclusions}
\label{conc}

Results from analyses performed with data of IT73 and IC79/IT73
have been presented. 
They can be considered the first comprehensive set of cosmic ray analyses
that incorporate detailed studies of the detector performance 
and systematics conducted during the construction years of
IceCube.
IceCube is currently taking data in the third year after its completion.
Upcoming analyses will yield refined results with smaller systematics
and larger statistics, and lead to new understanding about cosmic ray composition and sources.
Analysis of coincident events allows the use of the electromagnetic to TeV
muon component ratio whose outcome has not been explored before.
The goal is to provide a bridge between experiments operating at
lower energies around the knee and ultra-high energy detectors at EeV
energies.
At the moment, the insufficient knowledge of the ice's optical
properties does not allow a better estimate of the absolute average composition.

\section*{Acknowledgments}

This research is supported in part by the
U.S. National Science Foundation Grant NSF-ANT-1205809.






\begin{thebibliography}{00}

\bibitem{Blasi:2011fi} 
  P.~Blasi and E.~Amato,
  {\it JCAP} {\bf 1201}, 010 (2012),
  {\it arXiv}:1105.4521.

 \bibitem{Blasi:2011fl} 
  P.~Blasi and E.~Amato,
  {\it JCAP} {\bf 1201}, 011 (2012),
  {\it arXiv}:1105.4529.


\bibitem{Berezinsky:2007wf} 
  R. Aloisio, V. Berezinsky, A. Gazizov,
  {\it Astropart. Phys.} {\bf 39-40}, 129 (2012).

\bibitem{Kampert:2012mx} 
  K.~H.~Kampert and M.~Unger,
  {\it Astropart. Phys.} {\bf 35}, 660 (2012).

\bibitem{fyp} 
A.~Achterberg {\it et al.}  [IceCube Collaboration],
{\it Astropart. Phys.} {\bf 26}, 155 (2006).

\bibitem{ignacio} 
I.~Taboada for the IceCube Collaboration,
these proceedings.

\bibitem{DOM} 
R.~Abbasi {\it et al.}  [IceCube Collaboration],
{\it Nucl. Instrum. Methods A} {\bf 601}, 294 (2009).

\bibitem{IT} 
R.~Abbasi {\it et al.}  [IceCube Collaboration],
{\it arXiv}:1207.6326 (2012).

  \bibitem{snow} 
 IceCube Coll., paper 1106,  {\it Proc. of ICRC 2013}.
   
\bibitem{it26} 
R.~Abbasi {\it et al.}  [IceCube Collaboration],
{\it Astropart. Phys.} {\bf 44}, 40-58 (2013).

\bibitem{ic40} 
R.~Abbasi {\it et al.}  [IceCube Collaboration],
{\it Astropart. Phys.} {\bf 42}, 15-32 (2013).

 \bibitem{Abbasi:2010mf} 
  R.~Abbasi {\it et al.}  [IceCube Collaboration],
  {\it Astrophys. J.}  {\bf 718}, L194 (2010).

\bibitem{Abbasi:2011zka} 
  R.~Abbasi {\it et al.}  [IceCube Collaboration],
  {\it Astrophys. J.}  {\bf 746}, 33 (2012).
  
\bibitem{it73_1} 
M.~G.~Aartsen {\it et al.} [IceCube Collaboration], 
 	{\it arXiv}:1307.3795 (2013).

   \bibitem{Gaisser:2012zz} 
  T.~K.~Gaisser,
  {\it Astropart. Phys.}  {\bf 35}, 801 (2012).


\bibitem{sybill} 
E.~Ahn {\it et al.}, {\it Phys. Rev.} {\bf D 80}, 094003 (2009).

\bibitem{qgsjet} 
S.~Ostapchenko, {\it Nucl. Phys. Proc. Suppl.} {\bf B 151},  143-146 (2006).

\bibitem{nn}
 IceCube Coll., paper 0861,  {\it Proc. of ICRC 2013}.

\bibitem{seas}
 IceCube Coll., paper 0763,  {\it Proc. of ICRC 2013}.

\bibitem{dedx}
T.~Feusels {\it  et al.}, in {\it Proc. of ICRC 2009},
{\it arXiv}:0912.4668.

\bibitem{elbert}
J.~W.~Elbert, in {\it Proc. DUMAND Summer Workshop} (ed. A.
Roberts), vol. 2, p. 101 (1978).

\bibitem{ani}
P.~Desiati for the IceCube Collaboration, these proceedings.

\bibitem{itani}
M.~G.~Aartsen {\it et al.} [IceCube Collaboration], {\it Astrophys. J.}  {\bf 765}, 55 (2013).

      \bibitem{CG}
  A.~H.~Compton, I.~A.~Getting, 
 {\it  Phys. Rev.} {\bf 47} 817-821 (1935).

\bibitem{infill} 
 IceCube Coll., paper 0674,  {\it Proc. of ICRC 2013}.
   
\bibitem{inclined}
 IceCube Coll., paper 0973,  {\it Proc. of ICRC 2013}.

\end{thebibliography}



\end{document}